\documentclass[twocolumn,english,superscriptaddress,floatfix,longbibliography]{revtex4-1}
\usepackage[LGR,T1]{fontenc}
\usepackage[latin9]{inputenc}
\usepackage{amsmath}
\usepackage{amssymb}
\usepackage{graphicx}

\makeatletter

\DeclareRobustCommand{\greektext}{%
  \fontencoding{LGR}\selectfont\def\encodingdefault{LGR}}
\DeclareRobustCommand{\textgreek}[1]{\leavevmode{\greektext #1}}
\ProvideTextCommand{\~}{LGR}[1]{\char126#1}

\usepackage{braket}
\usepackage{upgreek}

\makeatother

\usepackage{babel}
\begin{document}
\title{Exciting terahertz magnons with amplitude modulated light: spin pumping,
squeezed states, symmetry breaking and pattern formation}
\author{Egor I. Kiselev}
\affiliation{Max-Planck-Institut für Physik komplexer Systeme, 01187 Dresden, Germany}
\author{Jonas F. Karcher}
\affiliation{Princeton University, Department of Electrical and Computer Engineering,
Princeton, New Jersey 08544, USA}
\author{Mark S. Rudner}
\affiliation{Department of Physics, University of Washington, Seattle, Washington
98195-1560, USA}
\author{Rembert Duine}
\affiliation{Institute for Theoretical Physics, Utrecht University, Princetonplein
5, 3584 CC Utrecht, The Netherlands}
\affiliation{Department of Applied Physics, Center for NanoMaterials (cNM) Eindhoven
University of Technology, P.O. Box 513, 5600 MB Eindhoven, The Netherlands}
\author{Netanel H. Lindner}
\affiliation{The Helen Diller Quantum Center, Technion, Haifa 3200003, Israel}
\begin{abstract}
We show how amplitude modulated, coherent high-frequency drives can
be used to access otherwise difficult to reach collective resonances
and off-resonantly induce parametric instabilities. In particular,
we demonstrate that difficult to access antiferromagnetic resonances
in the THz range can be parametrically excited with signals at optical
frequencies via a mechanism that we call Modulated Floquet Parametric
Driving (MFPD). We study spin pumping and the formation of entangled,
two-mode squeezed magnon pairs in anisotropic antiferromagnets under
MFPD. Furthermore, we show that MFPD induces transitions to symmetry
breaking steady-states in which dynamical spin patterns are formed
by resonant magnon pairs.
\end{abstract}
\maketitle

\section{Introduction}

\begin{figure}
\begin{centering}
\includegraphics[width=0.95\columnwidth]{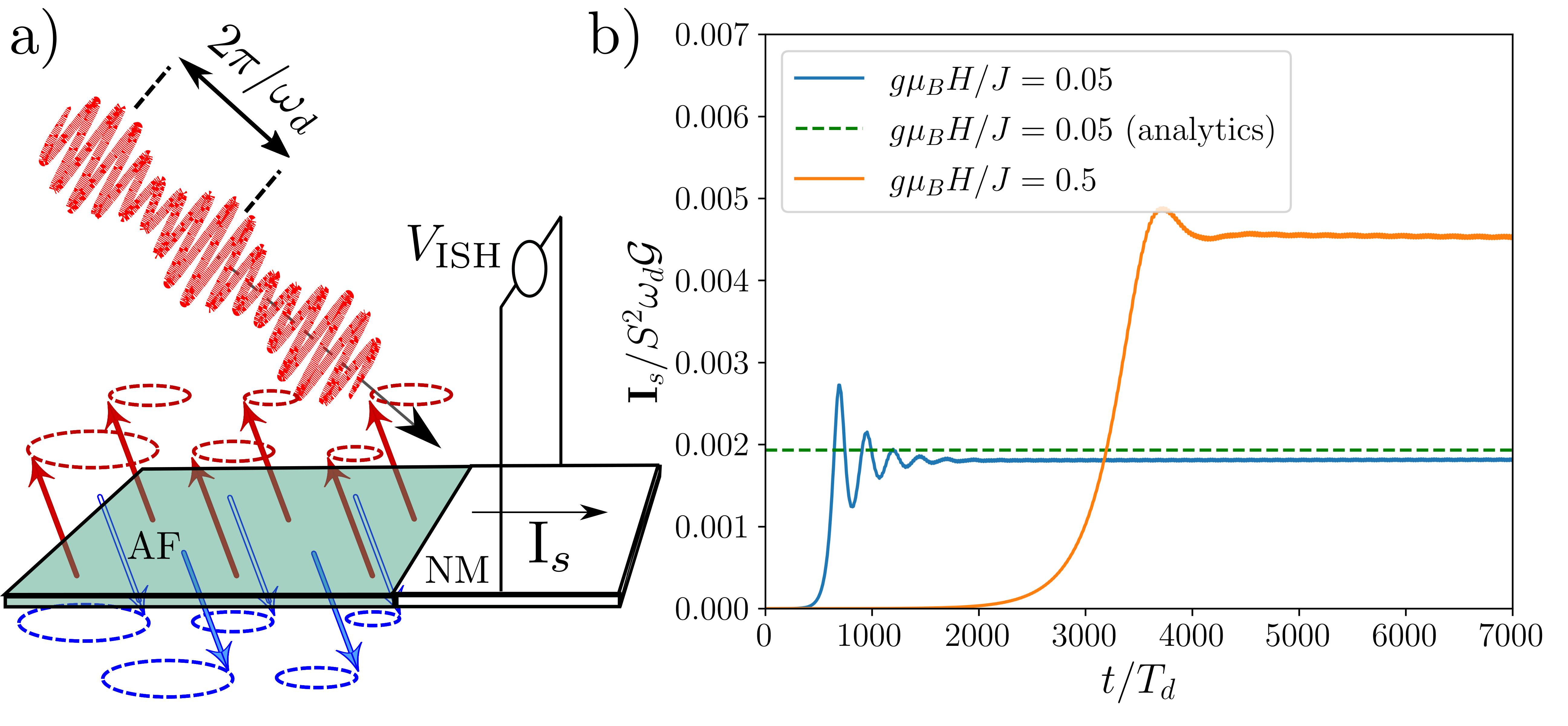}\caption{\label{fig:MFPD_principle_spin_pumping}\textbf{a)} Modulated Floquet
Parameteric Driving (MFPD) of an antiferromagnet (AF) -- a laser
beam modulated at a frequency $\omega_{d}$, is coherently driving
magnons which are in resonance with the modulation frequency. If the
AF is placed next to a normal metal (NM), a spin current $I_{s}$
is induced, which creats an inverse spin Hall voltage $V_{\mathrm{ISH}}$.
\textbf{b)} Spin current pumped from an MFPD driven AF for different
magnetic fields $\mathbf{H}$. We used the parameters $D_{z}=0.6J$,
$D_{x}=0.5J$, $\delta J=0.01$ and $S=1$. For small fields, the
result is well described by Eqs. (\ref{eq:delta_H_amplitudes}) and
(\ref{eq:spin_current_a_=00005Cdelta_a}).}
\par\end{centering}
\end{figure}

Antiferromagnets (AFs) host magnon modes with resonance frequencies
in the THz range, which, together with their immunity to stray magnetic
fields, makes AFs promising candidates for the realization of ultra-fast
devices for information processing and storing \citep{walowski2016perspective_thz_spintronics,jungwirth2018_directions_of_antiferromagnetic_spintronics,olejnik2018terahertz_memory_antiferromagnet,barman2021_magnonics_roadmap}.
Although considerable progress has been achieved in antiferromagnetic
spintronics over the past decade \citep{jungwirth2016_AF_spintronics_1,jungwirth2018_directions_of_antiferromagnetic_spintronics,vaidya2020subterahert_antiferromagnet_spin_pumping_exp,baltz2018_AF_spintronics_2,bilyk2025thz_AF,yang2024_sub_THz_hematite_manipulation_spin_orbit_torque,lebrun2020long_hematite_ultra_low_damping_1,el2023_hematite_ultra_low_damping_2,fritjofson2025coherent},
the control of THz magnons in AFs and their control remains a major
practical difficulty. On the other hand, the control of interacting
quantum matter by coherent light has recently attracted considerable
interest \citep{bloch2022strongly,basov2017towards,cavalleri2018photo,oka2019floquet,rudner2020floquet_review,oka2009photovoltaic,kitagawa2011transport,lindner2011floquet,fausti2011light,wang2013observation,lindner2013topological,mahmood2016selective,mciver2020light,esin2020floquet,esin2021electronic,zhou2023pseudospin,merboldt2025observation,choi2025observation}.
Here, we propose a method to coherently drive AF magnons in the THz
range, which does not employ THz sources. Instead, we suggest that
a coherent optical signal whose amplitude is modulated with the frequency
of the magnon resonance can be used to excite magnons parametrically
(see Fig. \ref{fig:MFPD_principle_spin_pumping} a)). We call this
new method -- which we applied to plasmons in previous publications
\citep{kiselev2024inducing,kiselev2023light} -- Modulated Floquet
Parametric Driving (MFPD). We note that signals with modulation frequencies
of up to $10$ THz can be created by interfering detuned (e.g. thermally)
laser sources -- a technique used in so called extraordinary acoustic
Raman spectroscopy \citep{wheaton2015_extraordinary_acoustic_raman,rubinsztein2016roadmap_structured_light}.

In the case of magnons, MFPD makes use of the modification of the
exchange constant $J$ in magnetic insulators by coherent light \citep{mentink2015_ultrafast_control_J,chaudhary2019orbital_floquet_engigeering_exchange,chaudhary2020_ligand_mediated_exchange_light,ron2020ultrafast_enhancement_of_exchange}.
However, compared to many previous proposals to manipulate magnetic
states via the light-induced modification of $J$ \citep{mentink2015_ultrafast_control_J,chaudhary2020_ligand_mediated_exchange_light,chaudhary2019orbital_floquet_engigeering_exchange,kumar2022floquet_Kitaev,arakawa2021floquet_SOC,vogl2022light_driven_magentic_transitions,rodriguez2022light_topo_magn},
MFPD is effective at more modest driving powers, since it couples
to a resonant collective mode -- the magnon -- which, due to its
high quality, can accumulate energy from the drive over many oscillation
cycles (see discussion section).

In the following, we elaborate on the fundamental mechanisms of MFPD
and explore four important applications: i) exciting THz magnons without
the need for THz sources, ii) inducing non-trivial symmetry breaking
steady states and the formation of dynamical spin patterns iii) pumping
spin from AFs to non-magnetic materials (see Fig. \ref{fig:MFPD_principle_spin_pumping})
-- an effect that has attracted a lot of interest recently \citep{cheng2014_antiferromagnet_spin_pumping,johansen2017spin_pumping_AF,vzelezny2018spin_pumping_torque_af,vaidya2020subterahert_antiferromagnet_spin_pumping_exp,lund2021spin_pumping_noncollinear_af,wang2021spin_pumping_DMI}
and could be achieved without THz radiation using MFPD, iv) creating
squeezed and entangled magnon states for applications in the emergent
field of quantum magnonics \citep{yuan2022quantum_magnonics,kamra2019antiferromagnetic,kamra2020magnon,romling2024quantum,guo2023magnon,li2019squeezed}. 

Recent experiments have demonstrated the extraordinarily high quality
of certain AF insulators, in particular the hematite \textgreek{a}-Fe$_{2}$O$_{3}$,
with damping factors as low as $\alpha\leq10^{-5}$ \citep{lebrun2020long_hematite_ultra_low_damping_1,el2023_hematite_ultra_low_damping_2}.
Thus the two main prerequisites of AF magnon MFPD -- ultra-high quality
AF resonances and THz modulated laser beams -- are within experimental
reach.

We start our discussion by introducing a model that captures the main
features of MFPD of magnons. A driven antiferromagnetic insulator
is modeled via the Hamiltonian
\begin{equation}
H=J\left(t\right)\sum_{\left\langle ij\right\rangle }\mathbf{S}_{i}\cdot\mathbf{S}_{ij}-D_{z}\sum_{i}S_{z,i}^{2}+D_{x}\sum_{i}S_{x,i}^{2}.\label{eq:toy_hamiltonian}
\end{equation}
Here, $D_{z}>0$ and $D_{x}>0$ are easy-axis and in-plane single-ion
anisotropy constants and $J\left(t\right)>0$ is the time dependent
exchange constant and $\left|\mathbf{S}\right|>1$. The first sum
extends over nearest neighbors. For the sake of simplicity, we consider
a 1D spin chain first, and generalize our results to two dimensions
later on. Because anisotropies break the rotational symmetry of spins,
magnon dispersions acquire a gap (see Fig. \ref{fig:magnon_dispersions}),
whose magnitude is typically in the THz or sub-THz range range (some
examples are the insulating hematite \textgreek{a}-Fe$_{2}$O$_{3}$
\citep{yang2024_sub_THz_hematite_manipulation_spin_orbit_torque},
MnF$_{2}$ \citep{vaidya2020subterahert_antiferromagnet_spin_pumping_exp}
and NiO \citep{rongione2023emission_THz_magnons_NiO}). 

The MFPD mechanism induces time dependence to $J\left(t\right)$ in
two conceptual steps. First, consider an oscillating electric field
with frequency $\Omega$ and ampltiude $E$ applied to the material.
This field, with a (non-dimensionalized) amplitude $\mathcal{E}=ea_{l}E/\left(\hbar\Omega\right)$,
where $a_{l}$ is the lattice constant, alters the exchange coupling:
$J\left(\mathcal{E}=0\right)\rightarrow J\left(\mathcal{E}\right)$
(see \ref{subsec:Toy-model-for_J(E)}). Next, suppose the driving
field is subjected to an amplitude modulation: $\mathcal{E}\left(t\right)=\bar{\mathcal{E}}\cos\left(\omega_{d}t\right)$.
If $\omega_{d}\ll\Omega$, the process is adiabatic, yielding
\begin{equation}
J\left(t\right)\approx\bar{J}+\delta J\cos\left(2\omega_{d}t\right)\label{eq:J(t)=00003DJ+=00005CdeltaJ}
\end{equation}
with $\delta J=\left(\mathcal{E}^{2}/4\right)\left.\partial^{2}J\left(\mathcal{E}\right)/\partial\mathcal{E}^{2}\right|_{\mathcal{E}=0}$.
A second order response of $J$ to $\mathcal{E}$ is expected generically,
as shown in the toy model of \ref{subsec:Toy-model-for_J(E)} and
we will use the simple result of Eq. (\ref{eq:J(t)=00003DJ+=00005CdeltaJ})
in what follows. Going beyond the mechanism presented in \ref{subsec:Toy-model-for_J(E)}
and its extensions to more realistic lattices \citep{anderson1950antiferromagnetism_superexchange,chaudhary2020_ligand_mediated_exchange_light},
microscopic resonances can be exploited to increase the effect of
the electromagnetic field on the exchange coupling \citep{ron2020ultrafast_enhancement_of_exchange}.

\section{Instabilities induced by time varying exchange}

Having introduced the notion of a time varying exchange coupling,
we now show how it can be used to parametrically pump magnons and
induce magnon instabilities. To investigate the magnon dynamics, we
use the Landau-Lifshitz-Gilbert (LLG) equation
\begin{equation}
\dot{\mathbf{S}}_{i}=\frac{1}{\hbar}\mathbf{S}_{i}\times\frac{\partial H}{\partial\mathbf{S}_{i}}+\alpha\mathbf{S}_{i}\times\dot{\mathbf{S}}_{i},\label{eq:LLG}
\end{equation}
where $\alpha$ is the Gilbert damping parameter. We first focus on
classical spins and discuss quantum effects in Sec. \ref{sec:Squeezing-and-quantum}.

\begin{figure}
\centering{}\includegraphics[width=0.95\columnwidth]{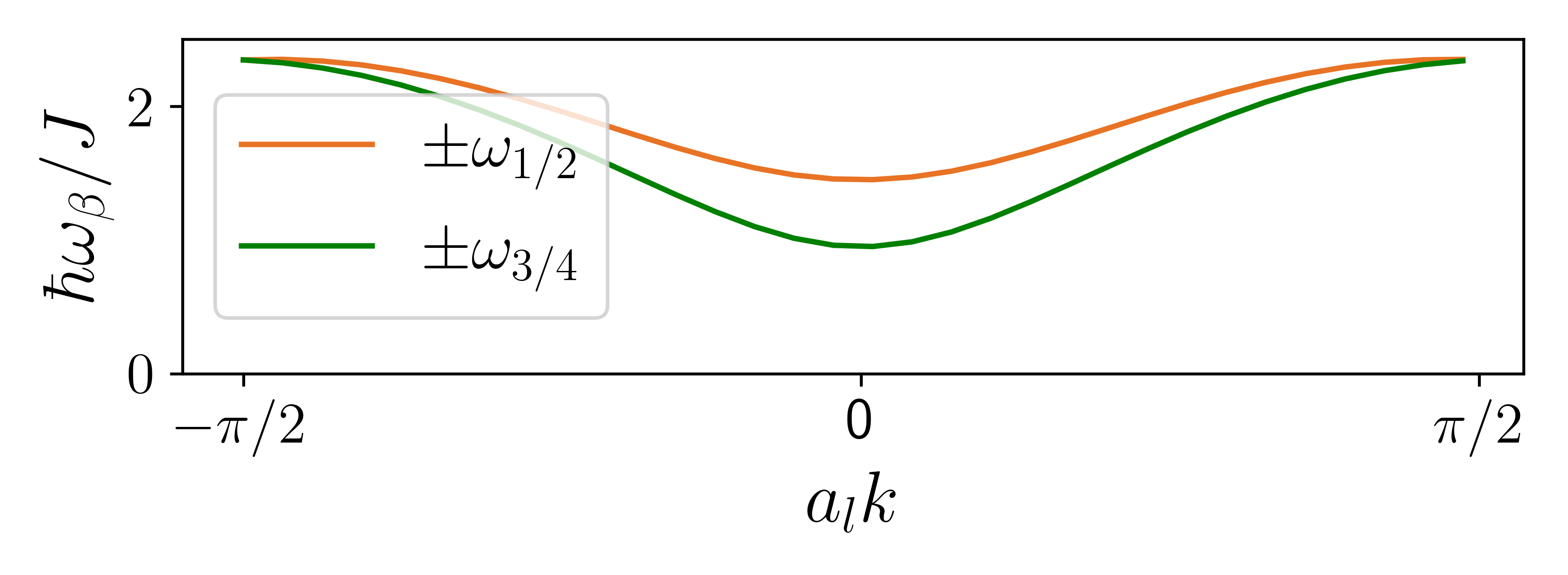}\caption{\label{fig:magnon_dispersions} Magnon dispersions according to Eq.
(\ref{fig:magnon_dispersions}). The branches $\omega_{1/2}$ and
$\omega_{3/4}$ are complex conjugate pairs. The dispersions are gapped
due to single-ion anisotropy. We used $D_{x}=0.3J$, $D_{y}=0.2J$
for the plot.}
\end{figure}

We begin with $\delta J=0$. Linearizing Eq. (\ref{eq:LLG}) around
the Néel state and anticipating wavelike solutions, we write for the
spin-up (A) and spin-down (B) sublattices: $\mathbf{S}_{j}^{\left(A/B\right)}=\left[\delta S_{x}^{\left(A/B\right)}\left(t\right)e^{ika_{l}j},\delta S_{y}^{\left(A/B\right)}\left(t\right)e^{ika_{l}j},\pm S\right]^{T}$,
where the plus/minus signs are chosen for the A/B sublattices, respectively
and $a_{l}$ is the lattice constant. To first order in the $\delta S$,
this gives the equation
\begin{equation}
\delta\dot{\vec{S}}=M\left(k\right)\delta\vec{S},\label{eq:magnon_dyn_eq}
\end{equation}
where $\delta\vec{S}=\left[\delta S_{x}^{\left(A\right)},\delta S_{y}^{\left(A\right)},\delta S_{x}^{\left(B\right)},\delta S_{y}^{\left(B\right)}\right]^{T}$
and $M\left(k\right)$ is a matrix determining the magnon dynamics
(see \ref{subsec:Linear_magnon_dynamics}). The magnon dispersions
$\omega_{\beta}\left(k\right)$ with $\beta\in\left\{ 1,2,3,4\right\} $
(see Fig. \ref{fig:magnon_dispersions}) are given by the eigenvalues
of $iM\left(k\right)$. Neglecting damping ($\alpha=0$), we find
\begin{align}
\omega_{1/2}\left(k\right) & =\pm S\sqrt{\left(2JL_{+,k}+D_{z}\right)\left(2JL_{-,k}+D_{x}+D_{z}\right)}\label{eq:magnon_dispersions}
\end{align}
where $L_{\pm,k}=\left[1\pm\cos\left(a_{l}k\right)\right]$, and $\omega_{3/4}\left(k\right)$
are obtained by interchanging $L_{+,k}$ and $L_{-,k}$.
Choosing a transformation matrix $P$ such that $P^{-1}MP=-i\mathrm{diag}\left\{ \omega_{1},\omega_{2},\omega_{3},\omega_{4}\right\} $,
the magnon eigenmodes are given by $\delta\vec{S}_{\mathrm{eig}}=P^{-1}\delta\vec{S}$,
and their dynamics follows from the equation
\begin{equation}
\delta\dot{S}_{\mathrm{eig},\beta}=-i\omega_{\beta}\delta S_{\mathrm{eig},\beta}.\label{eq:eigenmodes_EOM}
\end{equation}
Notice that $\omega_{1/2}\left(k\right)$ and $\omega_{3/4}\left(k\right)$
are complex conjugate frequency pairs.

The effect of a time dependent $J\left(t\right)$ is twofold (see
\ref{subsec:Linear_magnon_dynamics}). On the one hand, it will modify
the eigenvalues of Eq. (\ref{eq:eigenmodes_EOM}), introducing a time
dependent frequency shift. This effect is small for a small $\delta J$,
and we will neglect it. On the other hand, the eigenmodes $\delta S_{\mathrm{eig},1}$
and $\delta S_{\mathrm{eig},2}$, as well as $\delta S_{\mathrm{eig},3}$
and $\delta S_{\mathrm{eig},4}$ are coupled by a correction introduced
by the time dependent part of the exchange constant. This coupling
leads to a parametric resonance. Let us focus on the resonant case
of
\begin{equation}
\omega_{1/2}\left(k^{*}\right)=\pm\omega_{d},\label{eq:Resonance_and_k*}
\end{equation}
where $k^{*}$ is the wavenumber at which the eigenmodes $\delta S_{\mathrm{eig},1/2}$
is resonant with the modulation frequency $\omega_{d}$ (note that
$\omega_{\beta}\left(k\right)$ are non-degenerate, except for $ka=\pi/2$
-- we assume that the splitting between $\omega_{1/2}$ and $\omega_{3/4}$
is large enough, so that a resonance for one of the two pairs can
be selected). To first order in $\delta J$, we find that the set
of equations given in (\ref{eq:eigenmodes_EOM}) is modified, and,
for $\delta\dot{S}_{\mathrm{eig},1/2}$, reads
\begin{align}
\delta\dot{S}_{\mathrm{eig},1/2} & \approx\mp i\omega_{1}S_{\mathrm{eig},1/2}\mp i\omega_{1}\frac{\delta J}{J}C\cos\left(2\omega_{d}t\right)\delta S_{\mathrm{eig},2/1}.\label{eq:param_coupled_eq}
\end{align}
Here $C$ is a dimensionless quantity defined in \ref{subsec:Linear_magnon_dynamics}
that depends on $J$, $D_{x}$, $D_{z}$ and $k$. Eq. (\ref{eq:param_coupled_eq})
is derived in \ref{subsec:Linear_magnon_dynamics}. Using the slowly
varying envelope approximation (\ref{subsec:Linear_magnon_dynamics}),
we find the solutions
\begin{align}
\delta S_{\mathrm{eig},1/2} & =a_{1/2}e^{\frac{\delta JC}{2J}\omega_{d}t}e^{\pm i\omega_{d}t}\label{eq:growing_unstable_modes_no_damping}
\end{align}
The exponential growth in Eq. (\ref{eq:growing_unstable_modes_no_damping})
indicates a linear instability of the resonant magnons. Additionally,
there exist exponentially decaying solutions, which are not of interest
here. We include quantum-mechanical simulations of the instability
onset using TenPy \citep{hauschild2018tenpy1,hauschild2024tenpy2}
in \ref{subsec:Holstein-Primakoff-transformatio}.

So far, we did not include the Gilbert damping in our calculations.
This can be done perturbatively, and we find (\ref{subsec:Linear_magnon_dynamics})
that for a finite $\alpha$, the instability occurs if $\delta J$
exceeds a threshold value:
\begin{equation}
\frac{\delta J}{J}>\frac{1}{JSC}\frac{4\alpha\omega_{d}\hbar}{4J+D_{x}+2D_{z}}.\label{eq:=00005Cdelta_J_threshold}
\end{equation}
Thus, a small damping is preferable in order to keep the driving power
necessary to induce the instability low.

\section{Steady State and spin pumping}

In this section we study how the linear instability described above
is saturated by the nonlinearity of the spin system, resulting in
a steady state, in which the linear and nonlinear terms of the LLG
equation balance each other. We derive the magnon amplitudes in this
steady state analytically, focusing on the case of $k^{*}=0$, which
is realized for $\omega_{d}=\omega_{1}\left(k=0\right)$. We then
show that if a magnetic field along the $z$-axis is applied to the
system, the steady state can be used to pump spin into a neighboring
non-magnetic material. This provides one natural experimental test
for the parametric resonance and steady-state via the inverse spin
Hall effect (see Fig \ref{fig:MFPD_principle_spin_pumping} and the
Discussion section).

To find the magnon amplitudes in the steady states, we use the ansatz
\begin{align}
\delta S_{x}^{\left(A\right)} & =b\left(t\right)\sin\left(\omega_{d}t\right)\nonumber \\
\delta S_{y}^{\left(A\right)} & =a\left(t\right)\cos\left(\omega_{d}t\right)\nonumber \\
\delta S_{x}^{\left(B\right)} & =-b\left(t\right)\sin\left(\omega_{d}t\right)\nonumber \\
\delta S_{y}^{\left(B\right)} & =a\left(t\right)\cos\left(\omega_{d}t\right).\label{eq:steady_state_ansatz}
\end{align}
Inserting these formulae into the LLG equation (\ref{eq:LLG}) gives
a set of differential equations for the time-dependent amplitudes
$a\left(t\right)$, $b\left(t\right)$. Expanding these equations
to leading order in $\delta J$ and ignoring fast oscillations at
frequencies $2\omega_{d}$, $3\omega_{d}$, ..., we find the steady
state as a (non-trivial) fixed point with $\dot{a}=\dot{b}=0$ (see
\ref{subsec:Steady-state}), where
\begin{align}
a & =-b\sqrt{\frac{D_{x}+D_{z}}{4J+D_{z}}}\nonumber \\
b & =\pm\frac{2S\sqrt{\delta J}}{\sqrt{D_{x}+2D_{z}+4J}}.\label{eq:stead_state_amplitudes}
\end{align}

Next we add a Zeeman precession term of the from $-\mathbf{S}\times g\mu_{B}\mathbf{H}$
to Eq. (\ref{eq:LLG}). Solving for the fixed points of the amplitude
equations (see \ref{subsec:Steady-state}), we find the steady state
in the presence of a longitudinal magnetic field with $\mathbf{H}\propto\hat{\mathbf{e}}_{z}$.
To leading order in $\mathbf{H}$, the amplitudes of Eq. (\ref{eq:stead_state_amplitudes})
obtain corrections depending on the sublattice. For the $A$-sublattice,
we find $b\rightarrow b+\delta b_{H,A}$ and $a\rightarrow a+\delta a_{H,A}$.
For the $B$-sublattice amplitudes the corrections have the same magnitudes
but opposite signs. The expressions for $\delta b_{H,A}$, $\delta a_{H,A}$
are lengthy (see \ref{subsec:Steady-state}), but simplify for $J\gg D_{x},D_{z}$
and $D_{x}=D_{z}$. We then find
\begin{align}
\delta b_{H,A} & =g\mu_{B}H\frac{D_{z}b-\sqrt{8D_{z}J}a}{4JSD_{z}}\nonumber \\
\delta a_{H,A} & =g\mu_{B}H\frac{4Ja-\sqrt{8D_{z}J}b}{4JSD_{z}}.\label{eq:delta_H_amplitudes}
\end{align}
 If the antiferromagnet is put in proximity with a metal, the spin
current pumped into the metal is of the form \citep{cheng2014_antiferromagnet_spin_pumping,johansen2017spin_pumping_AF}
\begin{align}
\mathbf{I}_{s} & =\mathcal{G}\left(\mathbf{S}_{A}\times\dot{\mathbf{S}}_{A}+\mathbf{S}_{B}\times\dot{\mathbf{S}}_{B}\right)\label{eq:spin_current}
\end{align}
where the constant $\mathcal{G}$ is determined by the specific properties
of the magnet-to-metal interface. Averaging over one oscillation period
$2\pi/\omega_{d}$, only the $z$-component of Eq. (\ref{eq:spin_current})
contributes to a finite dc spin current. We obtain
\begin{equation}
I_{s,z}=2\omega_{d}\mathcal{G}\left(a\delta b_{H,A}+b\delta a_{H,A}\right).\label{eq:spin_current_a_=00005Cdelta_a}
\end{equation}
Notice that we assume the spin-pumping to be sufficiently weak that
the interface enhancement of the Gilbert damping is negligible. Fig.
\ref{fig:MFPD_principle_spin_pumping} b) shows the prediction of
Eq. (\ref{eq:spin_current_a_=00005Cdelta_a}) for the spin current
and the result obtained by solving Eq. (\ref{eq:LLG}) numerically.
For small $\mathbf{H}$ and $\delta J$, the two are in good agreement.

\section{Two dimensions: spin patterns in the steady state}

\begin{figure*}
\centering{}\includegraphics[width=0.75\paperwidth]{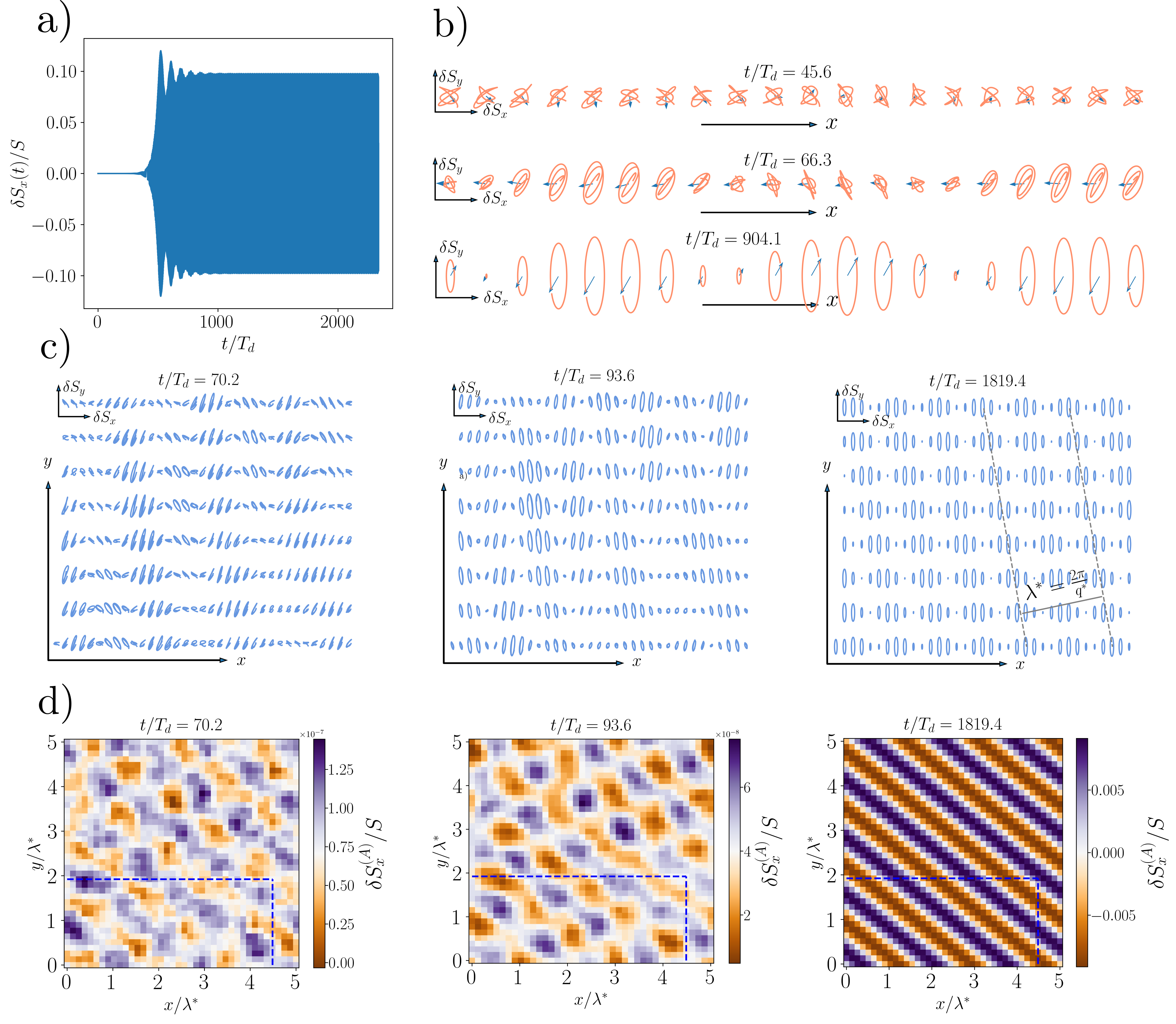}\caption{\label{fig:patterns}\textbf{a)} Spin oscillations in an MFPD driven
AF (1D chain). The oscillations of the spin component $S_{x}$ on
one of the two Néel sublattices are shown for driving above the threshold
(Eq. (\ref{eq:=00005Cdelta_J_threshold})). After a period of amplitude
oscillations, the spins enter a steady state with a constant amplitude.
$T_{d}=2\pi/\omega_{d}$ is the duration of one modulation cycle.
We used the parameters $D_{z}=0.2J$, $D_{x}=0.1J$, $\delta J=0.05J$,
$S=1$ and $\alpha=2.5\cdot10^{-3}$ throughout. \textbf{b)} Pattern
formation in a 1D spin chain. The figure shows the same simulation
as in a). When driven above threshold, spins begin to arrange themselves
in wave-like patterns with the characteristic wavelength $\lambda^{*}=2\pi/k^{*}$,
where $k^{*}$ is the wavenumber of resonant magnons, breaking the
translational invariance of the AF. \textbf{c)} and \textbf{d)} Pattern
formation in 2D for a square lattice. c) Shows individual spin trajectories
at different times $t$, whereas d) shows the magnitude of the spin
oscillation of $S_{x}$ on one of the Néel sublattices. Stripe patterns
breaking the translational and rotational symmetries of the system
emerge. The regions depicted in c) are marked with dashed blue lines
in d). All simulations were run on periodic domains.}
\end{figure*}
We now extend our results to two-dimensional systems and finite wavenumbers,
showing that symmetry breaking and dynamical spin patterns are induced
by MFPD, both in one and two dimensions for $k^{*}\neq0$. These patterns
share similarities with recently studied dynamical plasmon \citep{kiselev2023light,kiselev2024inducing}
and phonon states \citep{kaplan2025_phonons_faraday}. We perform
simulations of Eq. (\ref{eq:LLG}) for a 1D chain and for a two-dimensional
square lattice. A continuum model for Eq. (\ref{eq:LLG}) given in
\ref{subsec:Continuum_LLG} is used for the simulations ($k^{*}\ll1/a_{l}$).
The damping term is approximated as $\alpha\mathbf{S}_{i}\times\mathbf{\dot{S}}_{i}\approx\alpha\mathbf{S}_{i}\times\left(\mathbf{S}_{i}\times\partial H/\partial\mathbf{S}_{i}\right)/\hbar$.
This approximation can be obtained by iterating Eq. (\ref{eq:LLG})
in $\dot{\mathbf{S}}_{i}$ and ignoring higher order terms in $\alpha$. 

While for $k^{*}=0$ the LLG equations for the spin chain and square
lattice are identical up to prefactors stemming from the different
coordination numbers, the $k\neq0$ case is qualitatively distinct.
Here, the dispersions $\omega_{\beta}\left(\mathbf{k}\right)=\omega_{\beta}\left(k\right)$
are rotationally symmetric. Therefore, the resonance condition of
Eq. (\ref{eq:Resonance_and_k*}) is fulfilled for all wave vectors
with $\left|\mathbf{k}^{*}\right|=k^{*}$. As is typically the case
for parametrically driven non-linear systems \citep{cross1993patterns,muller1994model_patterns,chen1999_faraday_pattern_selection},
we expect that the system settles to a discrete set of wave vectors
$\left\{ \pm\mathbf{k}_{i}^{*}\right\} $ in the steady state, leading
to the emergence of a characteristic pattern that breaks the translational
and rotational symmetries of the system. 

In our simulations, the LLG equations for the chain and for the square
lattice (see \ref{subsec:Continuum_LLG}) are solved with the Dedalus
spectral solver \citep{burns2020dedalus}. For finite $k^{*}$, the
system, once driven beyond the instability threshold, after a series
of amplitude oscillations (Fig. \ref{fig:patterns} a)), converges
to a symmetry-breaking state corresponding to a magnonic standing
wave in one dimensions (Fig. \ref{fig:patterns} b)) and to a stripe
pattern in two dimensions (Fig. \ref{fig:patterns} c) and d)).These
states remain stable for as long as the simulations run (Fig. \ref{fig:patterns}
a)).

The striped state breaks the rotational and translational symmetries
of the system and the instability leading to its formation can be
seen as a dynamical phase transition \citep{kiselev2024inducing}. 

\section{Squeezing and quantum effects\label{sec:Squeezing-and-quantum}}

It is interesting to quantize the equation of motion (\ref{eq:param_coupled_eq}).
We focus on the AF chain in the regime of exponential growth, before
the onset of saturation and pattern formation. Promoting $\mathbf{S}_{i}$
to a quantum operator, we carry out a Holstein-Primakoff transformation
(\ref{subsec:Holstein-Primakoff-transformatio}), and find that the
creation and annihilation operators for magnons with momentum $k$
and frequency $\omega_{1}\left(k\right)$ are given by $\alpha_{k}^{\dagger}=\delta S_{\mathrm{eig},1,k}/\sqrt{N_{k}}$
and $\alpha_{k}=\delta S_{\mathrm{eig},1,k}^{\dagger}/\sqrt{N_{k}}$,
where $N_{k}$ is a normalizing factor (see \ref{subsec:Holstein-Primakoff-transformatio}
for commutation relations). Furthermore we find $\delta S_{\mathrm{eig},2,k}=\delta S_{\mathrm{eig},1,-k}^{\dagger}$,
such that the quantized version of the equations given in (\ref{eq:param_coupled_eq})
follow from the Hamiltionian
\begin{equation}
H=\sum_{k}\omega_{1,k}\left[\alpha_{k}^{\dagger}\alpha_{k}+\frac{\delta JC}{2}\cos\left(2\omega_{d}t\right)\left(\alpha_{k}^{\dagger}\alpha_{-k}^{\dagger}+\alpha_{k}\alpha_{-k}\right)\right].\label{eq:two_mode_squeezed_hamiltonian}
\end{equation}
The term proportional to $\delta J$ creates pairs of magnons with
opposite momenta, such that the overall momentum in the system is
conserved. Focusing on the resonant case of $k=k^{*}$, the time evolution
operator $U_{\mathrm{int}}\left(t,0\right)=\exp\left(-tH_{\mathrm{int}}/\hbar\right)$
with interaction picture Hamiltonian $H_{\mathrm{int}}=\frac{\delta J}{4J}C\left(k\right)\left(\alpha_{k^{*}}^{\dagger}\alpha_{-k^{*}}^{\dagger}+\alpha_{k^{*}}\alpha_{-k^{*}}\right)$
is a two-mode-squeezing operator with squeezing parameter $r=t\delta JC/4\hbar$
and phase $\theta=\pi/2$ (see Ref. \citep{gerry_knight_quantum_optics}
chapt. 7.7). The wave function in the basis $\ket{n_{k^{*}},n_{-k^{*}}}$,
where $n_{\pm k}$ are the magnon numbers reads
\begin{equation}
\ket{\psi\left(t\right)}=\frac{1}{\cosh\left(r\right)}\sum_{n=0}i^{n}\tanh^{n}\left(r\right)|n_{k^{*}},n_{-k^{*}}\rangle.
\end{equation}
This state is unfactorizable, and the magnon numbers for opposite
momenta are entangled due to the requirement of momentum conservation.
The number of magnons in each mode behaves as $\sinh^{2}\left(r\right)$,
which is in accordance with the classical result of Eq. (\ref{eq:growing_unstable_modes_no_damping}).
Maximal squeezing can be observed with the two quadrature operators
$X_{1}=2^{-3/2}e^{i\pi/4}(-i\alpha_{k^{*}}+\alpha_{k^{*}}^{\dagger}-i\alpha_{-k^{*}}+\alpha_{-k^{*}}^{\dagger})$
and $X_{2}=-2^{-3/2}e^{-i\pi/4}(i\alpha_{k^{*}}+\alpha_{k^{*}}^{\dagger}+i\alpha_{-k^{*}}+\alpha_{-k^{*}}^{\dagger})$,
where 
\begin{equation}
\left\langle X_{1/2}^{2}\right\rangle =e^{\mp2r}/4.
\end{equation}

\section{Discussion}

We have introduced MFPD -- a protocol to excite and control THz magnons
with amplitude modulated high frequency light. We have analyzed instabilities
and pattern formation in MFPD driven antiferromagnets, and pointed
at spin pumping and creating entangled, squeezed magnons applications. 

Amplitude modulated laser beams with tunable modulation frequencies
of up to 10 THz are achievable in practice \citep{rubinsztein2016roadmap_structured_light},
and have been used to excite sound modes of nanoparticles \citep{wheaton2015_extraordinary_acoustic_raman}.
This shows that MFPD can be used to excite magnons at otherwise difficult
to reach frequencies. In most previous works on light induced modifications
of the exchange constant \citep{mentink2015_ultrafast_control_J,mikhaylovskiy2015ultrafast_exchange_modification,chaudhary2019orbital_floquet_engigeering_exchange},
exceedingly high laser powers (with amplitudes of $10^{8}$ V/m and
more) were needed to achieve measurable effects. Indeed a simple estimation
shows that, using light with $\hbar\Omega=0.5\,\mathrm{eV}$, field
amplitudes of $10^{9}$ V/m are needed to achieve an effective amplitude
$\mathcal{E}=eaE/\left(\hbar\Omega\right)$ of order unity, which
would result in a strong modification of the exchange constant. In
our case, however, the required power is significantly lower due to
the high quality of antiferromagnetic resonances. Using Eq. (\ref{eq:=00005Cdelta_J_threshold})
and assuming $J\sim50\,\mathrm{meV}$, $D_{z}\sim D_{x}\sim0.1\,\mathrm{meV}$,
as well as a small, but not unrealistic damping of $\alpha=10^{-4}$
(damping lower than $10^{-5}$ has been measured in hematite \textgreek{a}-Fe$_{2}$O$_{3}$
\citep{lebrun2020long_hematite_ultra_low_damping_1}), we find that
the threshold exchange modulation is $\delta J/J\sim2\cdot10^{-5}$.
Since $\delta J\sim\sqrt{\mathcal{E}}$, this roughly corresponds
to a reduction of the required field amplitude by a factor of $2\cdot10^{2}$,
putting the required laser power into a much more accessible regime.
These estimates hold in an off-resonant regime. As outlined above,
MFPD does not depend on the microscopic details by which the exchange
is modulated. Coupling to intrinsic resonances can largely increase
the modification of $J$ by light \citep{ron2020ultrafast_enhancement_of_exchange},
which could help to operate MFPD at even lower powers. 

Finally, we estimate the inverse spin Hall voltage achievable with
MFPD at a two-dimensional interface between the light driven AF layer
and a normal metal. We use formulas (\ref{eq:stead_state_amplitudes}),
(\ref{eq:delta_H_amplitudes}) and (\ref{eq:spin_current_a_=00005Cdelta_a})
for a rough estimate of the spin current $I_{S}$. Let $\delta J$
be of the order of the threshold. With the above material parameters
and $g\mu_{B}H=0.1J$, we find $I_{S}\approx2.5\cdot10^{-6}\mathcal{G}\omega_{d}$.
We further estimate $\mathcal{G}\approx e/\text{Å}^{2}$ \citep{cheng2014_antiferromagnet_spin_pumping},
$\omega_{d}=1\,\mathrm{THz}$ and find $I_{S}=4\cdot10^{7}\,\mathrm{A/m^{2}}$.
With a typical value of $\theta_{\mathrm{SH}}=0.01$ for the spin
Hall angle \citep{miao2013inverse_Hall}, a conductivity of $\rho=100\,\upmu\Omega\mathrm{cm}$
\citep{miao2013inverse_Hall}, and a device width of $10\,\upmu\mathrm{m}$
with a film thickness of $5\,\mathrm{nm}$, we find a large inverse
spin Hall voltage of $\sim4\,\upmu\mathrm{V}$, implying that spin
pumping by MFPD should be readily observable.
\begin{acknowledgments}
We thank J. Schmalian for useful discussions. This project received
funding from the Horizon Europe Marie Skodowska-Curie Action program
under Grant Agreement 101155351. N.L. acknowledges support from ISF-MAFAT
Quantum Science and Technology Grant no. 2478/24. M.R. acknowledges
the Brown Investigator Award, a program of the Brown Science Foundation,
the University of Washington College of Arts and Sciences, and the
Kenneth K. Young Memorial Professorship for support.
\end{acknowledgments}

\newpage

\begin{widetext}

\section*{Supplementary material} 

\renewcommand{\thesection}{Supplement}%

\setcounter{figure}{0}
\renewcommand{\thefigure}{C\ \arabic{figure}}%

\setcounter{equation}{0}
\renewcommand{\theequation}{A\,\arabic{equation}}%

\subsection{Toy model for the modification of exchange by light\label{subsec:Toy-model-for_J(E)}}

In this supplement, we review a simple model for the light induced
modification of exchange in antiferromagnetic insulators \citep{mentink2015_ultrafast_control_J,chaudhary2019orbital_floquet_engigeering_exchange,chaudhary2020_ligand_mediated_exchange_light}.
Our starting point is the half filled 1D Hubbard model 
\begin{equation}
H=-t\sum_{\left\langle i,j\right\rangle }c_{i,\alpha}^{\dagger}c_{j,\alpha}+U\sum_{i}n_{i}n_{i}.\label{eq:1D_half_filled_Hubbard}
\end{equation}
Here, $c_{i,\alpha}^{\dagger}$ and $c_{j,\alpha}$ are operators
creating and annihilating electrons with spin $\alpha$ at sites $i$
and $j$, respectively, $n_{i}=\sum_{\alpha}c_{i,\alpha}^{\dagger}c_{i,\alpha}$
is the density operator at site $i$, $t$ is the hopping constant
and $U$ the on-site Coulomb repulsion. In the presence of electromagnetic
fields, the hopping constant is modified as prescribed by the Peierls
substitution
\begin{equation}
t\rightarrow te^{i\frac{e}{\hbar}\mathbf{a}_{l}\cdot\mathbf{A}\left(t\right)},
\end{equation}
where $\mathbf{A}\left(t\right)$ is the vector potential and $\mathbf{a}_{l}$
is the lattice vector. In the following, we will assume that the light
is polarized parallel to the chain, such that $\mathbf{A}\propto\mathbf{a}$.
Using the relation $\mathbf{E}=-\partial\mathbf{A}/\partial t$, we
write $A\left(t\right)=-E\cos\left(\Omega t\right)/\Omega$. For $U\gg t$,
a Schrieffer-Wolff transformation \citep{mentink2015_ultrafast_control_J}
yields the Heisenberg-Hamiltonian 
\begin{equation}
H=J\left(\mathcal{E}\right)\sum_{\left\langle i,j\right\rangle }\mathbf{S}_{i}\cdot\mathbf{S}_{j}
\end{equation}
with the exchange coupling 
\begin{equation}
J\left(\mathcal{E}\right)=\bar{J}\sum_{m=-\infty}^{\infty}\frac{J_{\left|m\right|}^{2}\left(\mathcal{E}\right)}{1+\frac{m\Omega}{U}},\label{eq:J_on_Epsilon}
\end{equation}
where $\mathcal{E}=ea_{l}E/\left(\hbar\Omega\right)$, $J_{\left|m\right|}\left(\mathcal{E}\right)$
are Bessel functions and $\bar{J}=4t^{2}/U$ is the unperturbed exchange
constant. Finally, we note, that the effect of the electromagnetic
field on the constants $D_{x}$ and $D_{z}$, in Eq. (\ref{eq:toy_hamiltonian})
is expected to be much smaller that the effect on the exchange, because
$D_{x}$ and $D_{z}$ are determined by on-site physics, such that
the electromagnetic field is experienced on a much shorter lengthscale.

The magnetic properties of many materials are shaped by ligand mediated
superexchange interactions \citep{anderson1950antiferromagnetism_superexchange}.
A careful analysis shows that the effect of an intermediate ligend
site on the light induced changes of the exchange coupling are small,
except in specific cases where on-site Coulomb repulsion and orbital
energy levels are fine tuned \citep{chaudhary2020_ligand_mediated_exchange_light}.

\subsection{Linear magnon dynamics and instabilities \label{subsec:Linear_magnon_dynamics}}

The equations of motion for the spin variables can be written as a
set of coupled first order differential equations (see Eq. (\ref{eq:magnon_dyn_eq})).
Explicitly, we have

\begin{equation}
\delta\dot{\vec{S}}=S\left[\begin{array}{cccc}
0 & -M_{A} & 0 & -M_{c}\\
M_{B} & 0 & M_{c} & 0\\
0 & M_{c} & 0 & M_{A}\\
-M_{c} & 0 & -M_{B} & 0
\end{array}\right]\delta\vec{S}\label{eq:matr_eq}
\end{equation}
with $M_{A}=2J\left(\text{\ensuremath{\mathcal{E}}}\right)+D_{z}$,
$M_{B}=2J\left(\text{\ensuremath{\mathcal{E}}}\right)+D_{z}+D_{x}$,
$M_{c}=2J\left(\text{\ensuremath{\mathcal{E}}}\right)\cos\left(ak\right)$.
For brevity, we suppress the dependency on $k$ in this supplement.
Let $P$ be the coordinate transform diagonalizing $M$, then
\begin{align*}
P^{-1}\delta\dot{\vec{S}} & =M_{\mathrm{diag}}P^{-1}\delta\vec{S},
\end{align*}
where $M_{\mathrm{diag}}=-i\mathrm{diag}\left\{ \omega_{1},\omega_{2},\omega_{3},\omega_{4}\right\} $.
The eigenfrequencies $\omega_{i}$ are given in Eq. (\ref{eq:magnon_dispersions}).
The eigenmodes $\delta S_{\mathrm{eig},i}$ are given by
\begin{align}
\delta S_{\mathrm{eig},1} & =\left(\delta S_{y}^{\left(A\right)}+\delta S_{y}^{\left(B\right)}\right)D_{1}+i\left(\delta S_{x}^{\left(A\right)}-\delta S_{x}^{\left(B\right)}\right)D_{4}\nonumber \\
\delta S_{\mathrm{eig},2} & =\left(\delta S_{y}^{\left(A\right)}+\delta S_{y}^{\left(B\right)}\right)D_{1}-i\left(\delta S_{x}^{\left(A\right)}-\delta S_{x}^{\left(B\right)}\right)D_{4}\nonumber \\
\delta S_{\mathrm{eig},3} & =\left(\delta S_{y}^{\left(A\right)}-\delta S_{y}^{\left(B\right)}\right)D_{2}+i\left(\delta S_{x}^{\left(A\right)}+\delta S_{x}^{\left(B\right)}\right)D_{3}\nonumber \\
\delta S_{\mathrm{eig},4} & =\left(\delta S_{y}^{\left(A\right)}-\delta S_{y}^{\left(B\right)}\right)D_{2}-i\left(\delta S_{x}^{\left(A\right)}+\delta S_{x}^{\left(B\right)}\right)D_{3}\label{eq:eigenmodes}
\end{align}
where the $D_{i}$ are defined as
\begin{align*}
D_{1} & =\sqrt{2\left[1+\cos(ak)\right]+D_{z}/J}\\
D_{2} & =\sqrt{2\left[1-\cos(ak)\right]+D_{z}/J}\\
D_{3} & =\sqrt{2\left[1+\cos(ak)\right]+D_{x}/J+D_{z}/J}\\
D_{4} & =\sqrt{2\left[1-\cos(ak)\right]+D_{x}/J+D_{z}/J}.
\end{align*}
When the time modulation of the exchange constant is switched on (see
Eq. (\ref{eq:J(t)=00003DJ+=00005CdeltaJ})), the eigenmodes $\delta S_{\mathrm{eig},1}$
and $\delta S_{\mathrm{eig},2}$, as well as $\delta S_{\mathrm{eig},3}$
and $\delta S_{\mathrm{eig},4}$ are coupled, since the $D_{i}$ depend
on $J$ in different ways. To first order in $\delta J/J$, and neglecting
terms which merely result in modifications of the $\omega_{i}$ and
$\delta S_{\mathrm{eig},i}$ of order $\delta J/J$, i.e. only keeping
terms which couple the eigenmodes and are responsible for the instability,
we arrive at Eq. (\ref{eq:param_coupled_eq}). 

A slowly varying amplitude approximation can now be used to solve
the Eqs. (\ref{eq:param_coupled_eq}). We make the ansatz
\begin{align*}
\delta S_{\mathrm{eig},1} & =a\left(t\right)e^{i\omega_{d}t}\\
\delta S_{\mathrm{eig},2} & =b\left(t\right)e^{-i\omega_{d}t},
\end{align*}
where we used that parametric resonance is expected for a wavenumber
$k^{*}$ where $\omega_{1}\left(ak^{*}\right)=\omega_{d}.$ Neglecting
higher harmonics in $\omega_{d}$, we find
\begin{align}
\dot{a}\left(t\right) & \approx i\omega_{d}\frac{\delta JC}{2}b\left(t\right)\nonumber \\
\dot{b}\left(t\right) & \approx-i\omega_{d}\frac{\delta JC}{2}a\left(t\right),\label{eq:s_v_env}
\end{align}
where 
\begin{align*}
C & =\frac{\partial\log D_{1}}{\partial J}-\frac{\partial\log D_{4}}{\partial J}.
\end{align*}
With the ansatz $a\left(t\right)=a_{0}e^{st}$, $b\left(t\right)=b_{0}e^{st}$,
we find
\begin{equation}
\left[\begin{array}{cc}
s & -i\omega_{d}\frac{\delta JC}{2}\\
i\omega_{d}\frac{\delta JC}{2} & s
\end{array}\right]\left[\begin{array}{c}
a_{0}\\
b_{0}
\end{array}\right]\approx0,\label{eq:slow_aplitudes_mateq}
\end{equation}
leading to
\begin{equation}
s=\pm\omega_{d}\frac{\delta JC}{2},\label{eq:growth_rate_no_damping}
\end{equation}
which gives the solutions of Eq. (\ref{eq:growing_unstable_modes_no_damping}). 

Gilbert damping, for a small $\alpha$, can be addressed perturbatively.
Writing the linearized damping term in Eq. (\ref{eq:LLG}) as
\[
\alpha\left[\begin{array}{cccc}
0 & -1 & 0 & 0\\
1 & 0 & 0 & 0\\
0 & 0 & 0 & -1\\
0 & 0 & 1 & 0
\end{array}\right]\left[\begin{array}{c}
\delta\dot{S}_{x}^{\left(A\right)}\\
\delta\dot{S}_{y}^{\left(A\right)}\\
\delta\dot{S}_{x}^{\left(B\right)}\\
\delta\dot{S}_{y}^{\left(B\right)}
\end{array}\right]=\alpha A\delta\dot{\vec{S}},
\]
adding this term to Eq. (\ref{eq:magnon_dyn_eq}), and expanding the
eigenvalues and eigenstates to leading order in $\alpha$, i.e. writing
$\omega_{\beta}=\omega_{\beta}^{\left(0\right)}+\alpha\omega_{\beta}^{\left(1\right)}$,
and similarly for $\delta\dot{\vec{S}}$, we find
\[
\sum_{\beta}\left[\mathbf{1}i\omega_{\beta}^{\left(0\right)}-M\right]\delta\vec{S}_{\mathrm{eig},\beta}^{\left(1\right)}=\sum_{\beta}\left[-\mathbf{1}\lambda_{\beta}^{\left(1\right)}+i\alpha A\omega_{\beta}^{\left(0\right)}\right]\delta\vec{S}_{\mathrm{eig},\beta}^{\left(0\right)}.
\]
With the usual argument that both $\delta\vec{S}_{\mathrm{eig},\beta}^{\left(1\right)}$
and $\delta\vec{S}_{\mathrm{eig},\beta}^{\left(1\right)}+\xi\delta\vec{S}_{\mathrm{eig},\beta}^{\left(0\right)}$
solve the above equation, and hence one can chose $\xi$ such that
$\left[\delta\vec{S}_{\mathrm{eig},\beta}^{\left(1\right)}\right]^{T}\cdot\delta\vec{S}_{\mathrm{eig},\beta}^{\left(0\right)}=0$,
we find
\begin{equation}
\omega_{\beta}^{\left(1\right)}=\alpha\left[\delta\vec{S}_{\mathrm{eig},\beta}^{\left(0\right)}\right]^{T}A\omega_{\beta}^{\left(0\right)}\delta\vec{S}_{\mathrm{eig},\beta}^{\left(0\right)}.\label{eq:damping_rate_perturb}
\end{equation}
Neglecting cross-terms of order $\mathcal{O}\left(\alpha\delta J\right)$,
Eq. (\ref{eq:slow_aplitudes_mateq}) becomes
\[
\left[\begin{array}{cc}
s-\lambda_{1}^{\left(1\right)} & -i\omega_{d}\frac{\delta JC}{2}\\
i\omega_{d}\frac{\delta JC}{2} & s-\lambda_{1}^{\left(1\right)}
\end{array}\right]\left[\begin{array}{c}
a_{0}\\
b_{0}
\end{array}\right]\approx0,
\]
such that Eq. (\ref{eq:growth_rate_no_damping}) is modified to
\begin{align*}
s & =-\frac{1}{S}\frac{2\alpha\omega_{d}^{2}}{4J+D_{x}+2D_{z}}\pm\frac{\omega_{d}\delta J}{2}C.
\end{align*}
This gives the instability threshold for $\delta J$ in Eq. (\ref{eq:=00005Cdelta_J_threshold}).

\subsection{Steady state\label{subsec:Steady-state}}

The amplitude equations resulting from inserting Eqs. (\ref{eq:steady_state_ansatz})
into the LLG, keeping $k=0$, and applying the approximations described
in the main text read
\begin{align}
b\omega_{d}= & \left(\frac{3D_{z}}{8S}+\frac{3J}{2S}+\epsilon\frac{\delta\tilde{J}}{S}\right)a^{3}+ab^{2}\left(\frac{D_{z}}{8S}+\frac{J}{2S}\right)\nonumber \\
\quad & -a\left(D_{z}S+4JS+2\epsilon S\delta\tilde{J}\right)\nonumber \\
-a\omega_{d} & =-\left(\frac{D_{x}}{8S}+\frac{D_{z}}{8S}\right)a^{2}b-b^{3}\left(\frac{3D_{x}}{8S}+\frac{3D_{z}}{8S}\right)\nonumber \\
 & \quad+bS\left(D_{x}+D_{z}\right).\label{eq:ampl_eqs}
\end{align}
We introduced a small scale $\epsilon\sim\delta J/J$ and wrote $\epsilon\delta\tilde{J}=\delta J$.
To zeroth order in $\epsilon$, we find the eigenvelue problem
\begin{align*}
a\left(D_{z}S+4JS\right) & =-b\omega_{d}\\
bS\left(D_{x}+D_{z}\right) & =-a\omega_{d},
\end{align*}
which is solved by $\omega_{d}=\omega_{1/2}=\pm S\sqrt{\left(4J+D_{z}\right)\left(D_{x}+D_{z}\right)}$
\begin{equation}
a=-b\sqrt{\frac{D_{x}+D_{z}}{4J+D_{z}}}.\label{eq:a_b_condition}
\end{equation}
To incorporate the nonlinear terms, we extend Eq. (\ref{eq:a_b_condition}):
\begin{equation}
a=-b\left(\sqrt{\frac{D_{x}+D_{z}}{4J+D_{z}}}+\epsilon X\right).\label{eq:a_b_cond_extend}
\end{equation}
Here $X$ is a correction to the eigenamplitudes. We insert Eq. (\ref{eq:a_b_cond_extend})
into Eqs. (\ref{eq:ampl_eqs}) and keep terms up to first order in
$\epsilon$. Solving the resulting equations for $b$ and $X$, we
find that to leading order in $\epsilon$, the amplitudes are given
by Eqs. (\ref{eq:stead_state_amplitudes}). 

Finally, in the presence of a magnetic field, ignoring small mixed
terms of the form $a^{2}\delta a_{H,B}$, $ab\delta b_{H,B}$, $\delta J\delta a_{H,B}$
and so on, we, from the LLG, we find
\begin{align*}
\delta a_{H,A} & =g\mu_{B}H\\
 & \times\frac{a\left(4J+D_{x}+D_{z}\right)-b\sqrt{\left(D_{x}+D_{z}\right)\left(4J+D_{z}\right)}}{4JSD_{x}}\\
\delta b_{H,A} & =g\mu_{B}H\frac{bD_{z}-a\sqrt{\left(D_{x}+D_{z}\right)\left(4J+D_{z}\right)}}{4JSD_{x}}
\end{align*}
and $\delta a_{H,B}=-\delta a_{H,A}$, $\delta b_{H,B}=-\delta b_{H,A}$.
It is important to note that these results are valid for $D_{x}\sim D_{z}$
since we assumed $2JD_{x}\gg D_{z}\delta J$, as well as $\delta a_{H,A/B}\ll a$
and $\delta b_{H,A/B}\ll b$.

\subsection{Quantum dynamics and Holstein-Primakoff transformation\label{subsec:Holstein-Primakoff-transformatio}}

A quantized picture of spin waves can be obtained with the Holstein-Primakoff
transformation of the original Heisenberg Hamiltonian. It is well
known, that, at the free magnon level, the spectrum coincides with
the one found by linearizing the classical equations of motion. Therefore,
we can restore the quantized dynamics from the classical results.
We focus on the one-dimensional chain. The Holstein-Primakoff transformation
on top of the Néel state prescribes the relations
\begin{align*}
\delta S_{x,k}^{\left(A\right)} & =\sqrt{\frac{S}{2}}\left(a_{k}+a_{-k}^{\dagger}\right),\ \delta S_{y,k}^{\left(A\right)}=-i\sqrt{\frac{S}{2}}\left(a_{k}-a_{-k}^{\dagger}\right)\\
\delta S_{x,k}^{\left(B\right)} & =\sqrt{\frac{S}{2}}\left(b_{k}+b_{-k}^{\dagger}\right),\ \delta S_{y,k}^{\left(B\right)}=i\sqrt{\frac{S}{2}}\left(b_{k}-b_{-k}^{\dagger}\right).
\end{align*}
From this, it can be shown that the operators $\delta S_{\mathrm{eig},1}$,
$\delta S_{\mathrm{eig},2}$ act as a pair of creation and annihilation
operators:
\begin{align}
\frac{\delta S_{\mathrm{eig},1,k}}{\sqrt{2SD_{1}\left(ak\right)D_{4}\left(ak\right)}} & =\alpha_{-k}^{\dagger}\nonumber \\
\frac{\delta S_{\mathrm{eig},2,k}}{\sqrt{2SD_{1}\left(ak\right)D_{4}\left(ak\right)}} & =\alpha_{k}.\nonumber \\
\left[\alpha_{k'},\alpha_{k}^{\dagger}\right] & =\delta\left(k-k'\right)\label{eq:eq:Eig_1_2_create_annihilate}
\end{align}
The equations of motion (\ref{eq:param_coupled_eq}), then lead to
the Hamiltonian (\ref{eq:two_mode_squeezed_hamiltonian}).

\begin{figure}
\begin{centering}
\includegraphics[width=0.6\textwidth]{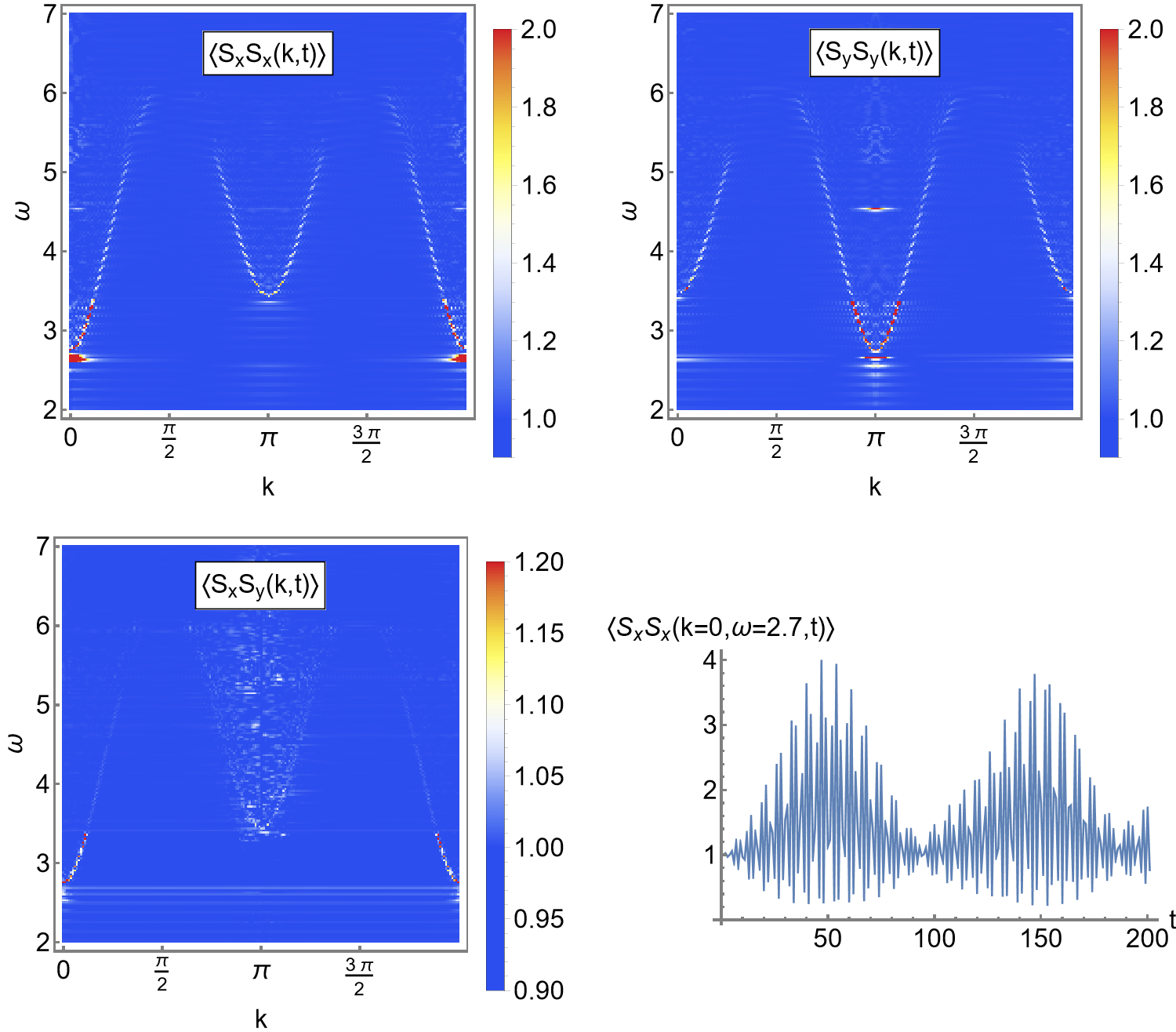}\caption{Quantum simulations of the instability of Hamiltonian (\ref{eq:toy_hamiltonian})
with a time dependent exchange constant (\ref{eq:J(t)=00003DJ+=00005CdeltaJ}). }
\par\end{centering}
\end{figure}

To illustrate the behavior of the quantum version of Eq. (\ref{eq:toy_hamiltonian}),
we performed numerical simulations of the driven interacting 1D chain.
We use TenPy \citep{hauschild2018tenpy1,hauschild2024tenpy2}. We
find the ground state $|\psi(0)>$ with iDMRG for the translationally
invariant system. The time evolvution of $|\psi(t)>$ is simulated
with iTEBD in steps of $dt=0.01$ for each omega. A bond dimension
of $\chi=120$ keeps the truncation error around $10^{-7}$. With
the obtained $|\psi(t)>$, we compute the correlation $C_{j}(t,\omega)=<\psi(t)|S(x=0)S(x=j)|\psi(t)>$
for j = -100,...,100. Panels (a), (b) and (c) show spatial Fourier
transform $C(q,t,\omega)$ of $C_{j}(t,\omega)$ for $t=200$. Panel
(d) shows a fixed q slice through (a) close to a resonance. We used
$S=5/2$, $D_{z}=0.1$, $D_{z}=-0.2$, and $\delta J=0.003$ in units
of $J=1$.

\subsection{Continuum version of the LLG in one and two dimensions\label{subsec:Continuum_LLG}}

Furthermore, for the simulations, we use the continuum versions of
Eq. (\ref{eq:LLG}), which are given in:
\begin{align}
\dot{\mathbf{S}}^{A/B} & =-\mathbf{S}^{A/B}\times\boldsymbol{\Omega}_{\mathrm{eff}}^{A/B}-\alpha\mathbf{S}^{A/B}\times\left(\mathbf{S}^{A/B}\times\boldsymbol{\Omega}_{\mathrm{eff}}^{A/B}\right),\label{eq:conti}
\end{align}
where we can approximate
\begin{align}
\hbar\boldsymbol{\Omega}_{\mathrm{eff}}^{A/B} & \approx-J\left(t\right)\left[2\mathbf{S}^{B/A}\left(x\right)+a^{2}\partial_{x}^{2}\mathbf{S}^{B/A}\left(x\right)\right]\nonumber \\
 & \quad+D_{z}S_{z}^{A/B}\hat{\mathbf{e}}_{z}-D_{x}S_{x}^{A/B}\hat{\mathbf{e}}_{x}\label{eq:conti_1d}
\end{align}
for a 1D chain and
\begin{align}
\hbar\boldsymbol{\Omega}_{\mathrm{eff}}^{A/B} & \approx-J\left(t\right)\left[4\mathbf{S}^{B/A}\left(\mathbf{x}\right)+a^{2}\nabla^{2}\mathbf{S}^{B/A}\left(\mathbf{x}\right)\right]\nonumber \\
 & \quad+D_{z}S_{z}^{A/B}\hat{\mathbf{e}}_{z}-D_{x}S_{x}^{A/B}\hat{\mathbf{e}}_{x}\label{eq:conti_2d}
\end{align}
for the Néel square lattice in two dimensions.

\end{widetext}


\begin{thebibliography}{10}

\bibitem{walowski2016perspective_thz_spintronics}
J.~Walowski and M.~M{\"u}nzenberg, \textit{Perspective: Ultrafast magnetism and
  thz spintronics}, Journal of Applied Physics \textbf{120} (2016).

\bibitem{jungwirth2018_directions_of_antiferromagnetic_spintronics}
T.~Jungwirth, J.~Sinova, A.~Manchon, X.~Marti, J.~Wunderlich, and C.~Felser,
  \textit{The multiple directions of antiferromagnetic spintronics}, Nature
  Physics \textbf{14}, 200 (2018).

\bibitem{olejnik2018terahertz_memory_antiferromagnet}
K.~Olejn{\'\i}k, T.~Seifert, Z.~Ka{\v{s}}par, V.~Nov{\'a}k, P.~Wadley, R.~P.
  Campion, M.~Baumgartner, P.~Gambardella, P.~N{\v{e}}mec, J.~Wunderlich
  \textit{et~al.}, \textit{Terahertz electrical writing speed in an
  antiferromagnetic memory}, Science advances \textbf{4}, eaar3566 (2018).

\bibitem{barman2021_magnonics_roadmap}
A.~Barman, G.~Gubbiotti, S.~Ladak, A.~O. Adeyeye, M.~Krawczyk, J.~Gr{\"a}fe,
  C.~Adelmann, S.~Cotofana, A.~Naeemi, V.~I. Vasyuchka \textit{et~al.},
  \textit{The 2021 magnonics roadmap}, Journal of Physics: Condensed Matter
  \textbf{33}, 413001 (2021).

\bibitem{jungwirth2016_AF_spintronics_1}
T.~Jungwirth, X.~Marti, P.~Wadley, and J.~Wunderlich, \textit{Antiferromagnetic
  spintronics}, Nature nanotechnology \textbf{11}, 231 (2016).

\bibitem{vaidya2020subterahert_antiferromagnet_spin_pumping_exp}
P.~Vaidya, S.~A. Morley, J.~van Tol, Y.~Liu, R.~Cheng, A.~Brataas, D.~Lederman,
  and E.~Del~Barco, \textit{Subterahertz spin pumping from an insulating
  antiferromagnet}, Science \textbf{368}, 160 (2020).

\bibitem{baltz2018_AF_spintronics_2}
V.~Baltz, A.~Manchon, M.~Tsoi, T.~Moriyama, T.~Ono, and Y.~Tserkovnyak,
  \textit{Antiferromagnetic spintronics}, Reviews of Modern Physics
  \textbf{90}, 015005 (2018).

\bibitem{bilyk2025thz_AF}
V.~Bilyk, R.~Dubrovin, A.~Zvezdin, A.~Kirilyuk, and A.~Kimel, \textit{Thz
  electric field control of spins in collinear antiferromagnet cr $ \_
  $\{$2$\}$ $ o $ \_ $\{$3$\}$ $}, arXiv preprint arXiv:2502.10181  (2025).

\bibitem{yang2024_sub_THz_hematite_manipulation_spin_orbit_torque}
D.~Yang, T.~Kim, K.~Lee, C.~Xu, Y.~Liu, F.~Wang, S.~Zhao, D.~Kumar, and
  H.~Yang, \textit{Spin-orbit torque manipulation of sub-terahertz magnons in
  antiferromagnetic $\alpha$-fe2o3}, Nature Communications \textbf{15}, 4046
  (2024).

\bibitem{lebrun2020long_hematite_ultra_low_damping_1}
R.~Lebrun, A.~Ross, O.~Gomonay, V.~Baltz, U.~Ebels, A.-L. Barra, A.~Qaiumzadeh,
  A.~Brataas, J.~Sinova, and M.~Kl{\"a}ui, \textit{Long-distance spin-transport
  across the morin phase transition up to room temperature in ultra-low damping
  single crystals of the antiferromagnet $\alpha$-fe2o3}, Nature communications
  \textbf{11}, 6332 (2020).

\bibitem{el2023_hematite_ultra_low_damping_2}
A.~El~Kanj, O.~Gomonay, I.~Boventer, P.~Bortolotti, V.~Cros, A.~Anane, and
  R.~Lebrun, \textit{Antiferromagnetic magnon spintronic based on nonreciprocal
  and nondegenerated ultra-fast spin-waves in the canted antiferromagnet
  $\alpha$-fe2o3}, Science Advances \textbf{9}, eadh1601 (2023).

\bibitem{fritjofson2025coherent}
G.~Fritjofson, A.~Regmi, J.~Hanson-Flores, J.~Michel, J.~Tang, F.~Yang,
  R.~Cheng, and E.~Del~Barco, \textit{Coherent spin pumping originated from
  sub-terahertz n$\backslash$'eel vector dynamics in easy plane
  $\{$$\backslash$alpha$\}$-fe2o3/pt}, arXiv preprint arXiv:2502.11281  (2025).

\bibitem{bloch2022strongly}
J.~Bloch, A.~Cavalleri, V.~Galitski, M.~Hafezi, and A.~Rubio, \textit{Strongly
  correlated electron--photon systems}, Nature \textbf{606}, 41 (2022).

\bibitem{basov2017towards}
D.~Basov, R.~Averitt, and D.~Hsieh, \textit{Towards properties on demand in
  quantum materials}, Nature materials \textbf{16}, 1077 (2017).

\bibitem{cavalleri2018photo}
A.~Cavalleri, \textit{Photo-induced superconductivity}, Contemporary Physics
  \textbf{59}, 31 (2018).

\bibitem{oka2019floquet}
T.~Oka and S.~Kitamura, \textit{Floquet engineering of quantum materials},
  Annual Review of Condensed Matter Physics \textbf{10}, 387 (2019).

\bibitem{rudner2020floquet_review}
M.~S. Rudner and N.~H. Lindner, \textit{Band structure engineering and
  non-equilibrium dynamics in floquet topological insulators}, Nature reviews
  physics \textbf{2}, 229 (2020).

\bibitem{oka2009photovoltaic}
T.~Oka and H.~Aoki, \textit{Photovoltaic hall effect in graphene}, Physical
  Review B \textbf{79}, 081406 (2009).

\bibitem{kitagawa2011transport}
T.~Kitagawa, T.~Oka, A.~Brataas, L.~Fu, and E.~Demler, \textit{Transport
  properties of nonequilibrium systems under the application of light:
  Photoinduced quantum hall insulators without landau levels}, Physical Review
  B \textbf{84}, 235108 (2011).

\bibitem{lindner2011floquet}
N.~H. Lindner, G.~Refael, and V.~Galitski, \textit{Floquet topological
  insulator in semiconductor quantum wells}, Nature Physics \textbf{7}, 490
  (2011).

\bibitem{fausti2011light}
D.~Fausti, R.~Tobey, N.~Dean, S.~Kaiser, A.~Dienst, M.~C. Hoffmann, S.~Pyon,
  T.~Takayama, H.~Takagi, and A.~Cavalleri, \textit{Light-induced
  superconductivity in a stripe-ordered cuprate}, science \textbf{331}, 189
  (2011).

\bibitem{wang2013observation}
Y.~Wang, H.~Steinberg, P.~Jarillo-Herrero, and N.~Gedik, \textit{Observation of
  floquet-bloch states on the surface of a topological insulator}, Science
  \textbf{342}, 453 (2013).

\bibitem{lindner2013topological}
N.~H. Lindner, D.~L. Bergman, G.~Refael, and V.~Galitski, \textit{Topological
  floquet spectrum in three dimensions via a two-photon resonance}, Physical
  Review B \textbf{87}, 235131 (2013).

\bibitem{mahmood2016selective}
F.~Mahmood, C.-K. Chan, Z.~Alpichshev, D.~Gardner, Y.~Lee, P.~A. Lee, and
  N.~Gedik, \textit{Selective scattering between floquet--bloch and volkov
  states in a topological insulator}, Nature Physics \textbf{12}, 306 (2016).

\bibitem{mciver2020light}
J.~W. McIver, B.~Schulte, F.-U. Stein, T.~Matsuyama, G.~Jotzu, G.~Meier, and
  A.~Cavalleri, \textit{Light-induced anomalous hall effect in graphene},
  Nature physics \textbf{16}, 38 (2020).

\bibitem{esin2020floquet}
I.~Esin, M.~S. Rudner, and N.~H. Lindner, \textit{Floquet metal-to-insulator
  phase transitions in semiconductor nanowires}, Science advances \textbf{6},
  eaay4922 (2020).

\bibitem{esin2021electronic}
I.~Esin, G.~K. Gupta, E.~Berg, M.~S. Rudner, and N.~H. Lindner,
  \textit{Electronic floquet gyro-liquid crystal}, Nature communications
  \textbf{12}, 5299 (2021).

\bibitem{zhou2023pseudospin}
S.~Zhou, C.~Bao, B.~Fan, H.~Zhou, Q.~Gao, H.~Zhong, T.~Lin, H.~Liu, P.~Yu,
  P.~Tang \textit{et~al.}, \textit{Pseudospin-selective floquet band
  engineering in black phosphorus}, Nature \textbf{614}, 75 (2023).

\bibitem{merboldt2025observation}
M.~Merboldt, M.~Schüler, D.~Schmitt, J.~P. Bange, W.~Bennecke, K.~Gadge,
  K.~Pierz, H.~W. Schumacher, M.~Davood, D.~Steil, S.~R. Manmana, M.~A. Sentef,
  M.~Reutzel, and S.~Mathias, \textit{Observation of floquet states in
  graphene}, Nature Physics  (2025).

\bibitem{choi2025observation}
D.~Choi, M.~Mogi, U.~De~Giovannini, D.~Azoury, B.~Lv, Y.~Su, H.~H{\"u}bener,
  A.~Rubio, and N.~Gedik, \textit{Observation of floquet--bloch states in
  monolayer graphene}, Nature Physics (1--6) (2025).

\bibitem{kiselev2024inducing}
E.~I. Kiselev, M.~S. Rudner, and N.~H. Lindner, \textit{Inducing exceptional
  points, enhancing plasmon quality and creating correlated plasmon states with
  modulated floquet parametric driving}, Nature Communications \textbf{15},
  9914 (2024).

\bibitem{kiselev2023light}
E.~I. Kiselev, Y.~Pan, and N.~H. Lindner, \textit{Light-controlled terahertz
  plasmonic time-varying media: Momentum gaps, entangled plasmon pairs, and
  pulse-induced time reversal}, Physical Review B \textbf{110}, L241411 (2024).

\bibitem{wheaton2015_extraordinary_acoustic_raman}
S.~Wheaton, R.~M. Gelfand, and R.~Gordon, \textit{Probing the raman-active
  acoustic vibrations of nanoparticles with extraordinary spectral resolution},
  Nature Photonics \textbf{9}, 68 (2015).

\bibitem{rubinsztein2016roadmap_structured_light}
H.~Rubinsztein-Dunlop, A.~Forbes, M.~V. Berry, M.~R. Dennis, D.~L. Andrews,
  M.~Mansuripur, C.~Denz, C.~Alpmann, P.~Banzer, T.~Bauer \textit{et~al.},
  \textit{Roadmap on structured light}, Journal of Optics \textbf{19}, 013001
  (2016).

\bibitem{mentink2015_ultrafast_control_J}
J.~Mentink, K.~Balzer, and M.~Eckstein, \textit{Ultrafast and reversible
  control of the exchange interaction in mott insulators}, Nature
  communications \textbf{6}, 6708 (2015).

\bibitem{chaudhary2019orbital_floquet_engigeering_exchange}
S.~Chaudhary, D.~Hsieh, and G.~Refael, \textit{Orbital floquet engineering of
  exchange interactions in magnetic materials}, Physical Review B \textbf{100},
  220403 (2019).

\bibitem{chaudhary2020_ligand_mediated_exchange_light}
S.~Chaudhary, A.~Ron, D.~Hsieh, and G.~Refael, \textit{Controlling
  ligand-mediated exchange interactions in periodically driven magnetic
  materials}, arXiv preprint arXiv:2009.00813  (2020).

\bibitem{ron2020ultrafast_enhancement_of_exchange}
A.~Ron, S.~Chaudhary, G.~Zhang, H.~Ning, E.~Zoghlin, S.~Wilson, R.~Averitt,
  G.~Refael, and D.~Hsieh, \textit{Ultrafast enhancement of ferromagnetic spin
  exchange induced by ligand-to-metal charge transfer}, Physical Review Letters
  \textbf{125}, 197203 (2020).

\bibitem{kumar2022floquet_Kitaev}
U.~Kumar, S.~Banerjee, and S.-Z. Lin, \textit{Floquet engineering of kitaev
  quantum magnets}, Communications Physics \textbf{5}, 157 (2022).

\bibitem{arakawa2021floquet_SOC}
N.~Arakawa and K.~Yonemitsu, \textit{Floquet engineering of mott insulators
  with strong spin-orbit coupling}, Physical Review B \textbf{103}, L100408
  (2021).

\bibitem{vogl2022light_driven_magentic_transitions}
M.~Vogl, S.~Chaudhary, and G.~A. Fiete, \textit{Light driven magnetic
  transitions in transition metal dichalcogenide heterobilayers}, Journal of
  Physics: Condensed Matter \textbf{35}, 095801 (2022).

\bibitem{rodriguez2022light_topo_magn}
M.~Rodriguez-Vega, Z.-X. Lin, A.~Leonardo, A.~Ernst, M.~G. Vergniory, and G.~A.
  Fiete, \textit{Light-driven topological and magnetic phase transitions in
  thin layer antiferromagnets}, The Journal of Physical Chemistry Letters
  \textbf{13}, 4152 (2022).

\bibitem{cheng2014_antiferromagnet_spin_pumping}
R.~Cheng, J.~Xiao, Q.~Niu, and A.~Brataas, \textit{Spin pumping and
  spin-transfer torques in antiferromagnets}, Physical review letters
  \textbf{113}, 057601 (2014).

\bibitem{johansen2017spin_pumping_AF}
{\O}.~Johansen and A.~Brataas, \textit{Spin pumping and inverse spin hall
  voltages from dynamical antiferromagnets}, Physical Review B \textbf{95},
  220408 (2017).

\bibitem{vzelezny2018spin_pumping_torque_af}
J.~{\v{Z}}elezn{\`y}, P.~Wadley, K.~Olejn{\'\i}k, A.~Hoffmann, and H.~Ohno,
  \textit{Spin transport and spin torque in antiferromagnetic devices}, Nature
  Physics \textbf{14}, 220 (2018).

\bibitem{lund2021spin_pumping_noncollinear_af}
M.~A. Lund, A.~Salimath, and K.~M. Hals, \textit{Spin pumping in noncollinear
  antiferromagnets}, Physical Review B \textbf{104}, 174424 (2021).

\bibitem{wang2021spin_pumping_DMI}
H.~Wang, Y.~Xiao, M.~Guo, E.~Lee-Wong, G.~Q. Yan, R.~Cheng, and C.~R. Du,
  \textit{Spin pumping of an easy-plane antiferromagnet enhanced by
  dzyaloshinskii--moriya interaction}, Physical Review Letters \textbf{127},
  117202 (2021).

\bibitem{yuan2022quantum_magnonics}
H.~Yuan, Y.~Cao, A.~Kamra, R.~A. Duine, and P.~Yan, \textit{Quantum magnonics:
  When magnon spintronics meets quantum information science}, Physics Reports
  \textbf{965}, 1 (2022).

\bibitem{kamra2019antiferromagnetic}
A.~Kamra, E.~Thingstad, G.~Rastelli, R.~A. Duine, A.~Brataas, W.~Belzig, and
  A.~Sudb{\o}, \textit{Antiferromagnetic magnons as highly squeezed fock states
  underlying quantum correlations}, Physical Review B \textbf{100}, 174407
  (2019).

\bibitem{kamra2020magnon}
A.~Kamra, W.~Belzig, and A.~Brataas, \textit{Magnon-squeezing as a niche of
  quantum magnonics}, Applied Physics Letters \textbf{117} (2020).

\bibitem{romling2024quantum}
A.-L.~E. R{\"o}mling and A.~Kamra, \textit{Quantum sensing of antiferromagnetic
  magnon two-mode squeezed vacuum}, Physical Review B \textbf{109}, 174410
  (2024).

\bibitem{guo2023magnon}
Q.~Guo, J.~Cheng, H.~Tan, and J.~Li, \textit{Magnon squeezing by two-tone
  driving of a qubit in cavity-magnon-qubit systems}, Physical Review A
  \textbf{108}, 063703 (2023).

\bibitem{li2019squeezed}
J.~Li, S.-Y. Zhu, and G.~Agarwal, \textit{Squeezed states of magnons and
  phonons in cavity magnomechanics}, Physical Review A \textbf{99}, 021801
  (2019).

\bibitem{rongione2023emission_THz_magnons_NiO}
E.~Rongione, O.~Gueckstock, M.~Mattern, O.~Gomonay, H.~Meer, C.~Schmitt,
  R.~Ramos, T.~Kikkawa, M.~Mi{\v{c}}ica, E.~Saitoh \textit{et~al.},
  \textit{Emission of coherent thz magnons in an antiferromagnetic insulator
  triggered by ultrafast spin--phonon interactions}, Nature communications
  \textbf{14}, 1818 (2023).

\bibitem{anderson1950antiferromagnetism_superexchange}
P.~W. Anderson, \textit{Antiferromagnetism. theory of superexchange
  interaction}, Physical Review \textbf{79}, 350 (1950).

\bibitem{hauschild2018tenpy1}
J.~Hauschild and F.~Pollmann, \textit{Efficient numerical simulations with
  tensor networks: Tensor network python (tenpy)}, SciPost Physics Lecture
  Notes (005) (2018).

\bibitem{hauschild2024tenpy2}
J.~Hauschild, J.~Unfried, S.~Anand, B.~Andrews, M.~Bintz, U.~Borla, S.~Divic,
  M.~Drescher, J.~Geiger, M.~Hefel \textit{et~al.}, \textit{Tensor network
  python (tenpy) version 1}, arXiv preprint arXiv:2408.02010  (2024).

\bibitem{kaplan2025_phonons_faraday}
D.~Kaplan, P.~A. Volkov, A.~Chakraborty, Z.~Zhuang, and P.~Chandra,
  \textit{Tunable spatiotemporal orders in driven insulators}, Physical Review
  Letters \textbf{134}, 066902 (2025).

\bibitem{cross1993patterns}
M.~C. Cross and P.~C. Hohenberg, \textit{Pattern formation outside of
  equilibrium}, Reviews of modern physics \textbf{65}, 851 (1993).

\bibitem{muller1994model_patterns}
H.~W. M{\"u}ller, \textit{Model equations for two-dimensional quasipatterns},
  Physical Review E \textbf{49}, 1273 (1994).

\bibitem{chen1999_faraday_pattern_selection}
P.~Chen and J.~Vinals, \textit{Amplitude equation and pattern selection in
  faraday waves}, Physical Review E \textbf{60}, 559 (1999).

\bibitem{burns2020dedalus}
K.~J. Burns, G.~M. Vasil, J.~S. Oishi, D.~Lecoanet, and B.~P. Brown,
  \textit{Dedalus: A flexible framework for numerical simulations with spectral
  methods}, Physical Review Research \textbf{2}, 023068 (2020).

\bibitem{gerry_knight_quantum_optics}
C.~C. Gerry and P.~L. Knight, \textit{Introductory quantum optics}, Cambridge
  university press (2023).

\bibitem{mikhaylovskiy2015ultrafast_exchange_modification}
R.~Mikhaylovskiy, E.~Hendry, A.~Secchi, J.~H. Mentink, M.~Eckstein, A.~Wu,
  R.~Pisarev, V.~Kruglyak, M.~Katsnelson, T.~Rasing \textit{et~al.},
  \textit{Ultrafast optical modification of exchange interactions in iron
  oxides}, Nature communications \textbf{6}, 8190 (2015).

\bibitem{miao2013inverse_Hall}
B.~Miao, S.~Huang, D.~Qu, and C.~Chien, \textit{Inverse spin hall effect in a
  ferromagnetic metal}, Physical review letters \textbf{111}, 066602 (2013).

\end{thebibliography}
\end{document}